\documentclass[prb,superscriptaddress,preprint,showpacs]{revtex4}
\usepackage{graphicx}
\usepackage{dcolumn}
\usepackage{bm}

\newcommand{\sigmapi}{\ensuremath{\sigma\rightarrow\pi}}
\newcommand{\sigmasigma}{\ensuremath{\sigma\rightarrow\sigma}}
\newcommand{\unpo}{U$_{0.75}$Np$_{0.25}$O$_{2}$}
\renewcommand{\vec}{\bm}
\newcommand{\miller}[3]{\ensuremath{(#1, #2, #3)}}
\renewcommand{\deg}{\ensuremath{^{\circ}}}

\DeclareGraphicsExtensions{.jpg,.eps,.png}

\begin{document}

\preprint{SBW/UNpO2/Figures V.1}
\title{Resonant X-ray Scattering Study of Magnetic and 
Electric-quadrupole
Order in U$_{0.75}$Np$_{0.25}$O$_2$}

\author{S. B. Wilkins}
\affiliation{European Commission, Joint Research Centre, Institute for
   Transuranium Elements, Postfach 2340, Karlsruhe, D-76125 Germany\\}
\affiliation{European Synchrotron Radiation Facility,
   Bo\^\i te Postal 220, F-38043 Grenoble CEDEX, France\\}
   \author{J. A. Paix\~{a}o}
\affiliation{Physics Department, University of Coimbra,
   Coimbra, 3004--516 Portugal\\}
\author{R. Caciuffo}
\affiliation{European Commission, Joint Research Centre, Institute for
   Transuranium Elements, Postfach 2340, Karlsruhe, D-76125 Germany\\}
\affiliation{Dipartimento di Fisica ed Ingegneria dei Materiali e del 
Territorio, Universit\`{a} Politecnica delle Marche, I-60131 Ancona, 
Italy\\}
\author{P. Javorsky}
\author{F. Wastin}
\author{J. Rebizant}
\affiliation{European Commission, Joint Research Centre, Institute for
   Transuranium Elements, Postfach 2340, Karlsruhe, D-76125 Germany\\}
\author{C. Detlefs}
\affiliation{European Synchrotron Radiation Facility,
   Bo\^\i te Postal 220, F-38043 Grenoble CEDEX, France\\}
\author{N. Bernhoeft}
\affiliation{D\'{e}partment de la Recherche Fondamentale sur la
   Mati\`{e}re Condens\'{e}e, CEA, F-38054 Grenoble CEDEX, France\\}
\author{P.Santini}
\affiliation{Dipartimento di Fisica, Universit\`{a} di Parma,Viale 
Parco delle Scienze, I-43100, Parma, Italy\\}
\author{G. H. Lander}
\affiliation{European Commission, Joint Research Centre, Institute for
   Transuranium Elements, Postfach 2340, Karlsruhe, D-76125 Germany\\}

\date{\today}

\begin{abstract}
We have used element specific X-ray resonant scattering to investigate the
M edge resonances in a single crystal of \unpo.
Earlier neutron diffraction and M\"{o}ssbauer studies had shown the
presence of long-range AF order below T$_{N}$ = 19~K, with sizeable magnetic 
moment both on the U and the Np ions. RXS results confirm the presence of an 
ordered dipole magnetic moment on the Np ions, but also reveal the presence of
an anti-ferro arrangement of the electric quadrupole moments on both U and Np ions, with the same 
propagation vector that defines the magnetic structure. From the azimuthal dependence of the intensities we are able to determine the exact configuration of the quadrupolar ordering.
The intensities of Bragg peaks associated with magnetic dipole 
and electric quadrupole order have different temperature dependences.
On cooling below T$_{N}$, the magnetic dipole order develops faster on the
uranium ions, with magnetic order on the Np ions increasing at a lower
rate. At the same temperature, quadrupolar order on both the U
and Np ions occurs along with an internal Jahn-Teller lattice distortion.

\end{abstract}

\maketitle

\section{Introduction}

Actinide dioxides, $An\mathrm{O_{2}}$, have fascinating physical
properties whose understanding is still a challenge, despite half a century of
extensive investigations carried out by both theoreticians and experimentalists.
The complexity of their physical behavior stems from the subtle interplay
between crystal field (CF), superexchange interactions and
electron-phonon coupling, which results in peculiar multi-polar
ordered-phase diagrams.

At room temperature
$An\mathrm{O_2}$ systems have
the $fcc$
crystal structure of $\mathrm{CaF_2}$, with space group $Fm\overline{3}m$ and
tetravalent $An^{4+}$ ions.
Ordered magnetism below $T_{N}=30.8\;K$ was suggested in 
$\mathrm{UO_2}$
by the presence of a
large anomaly in the temperature dependance of the heat
capacity\cite{Jones1952}, and
confirmed by magnetic susceptibility\cite{Arrot1957}
and neutron diffraction\cite{Willis1965,Frazer1965} measurements.
Neutron spectroscopy was soon after applied to elucidate collective
excitations, vibrational and magnetic, in the low temperature
phase\cite{Cowley1967}. Evidence of a strong spin-lattice interaction
prompted a theory introducing the concepts of Jahn-Teller (J-T) 
effects
and electric quadrupole interactions in the actinide
materials\cite{Allen1968}. Some years later, neutron diffraction
actually revealed an internal J-T distortion of the oxygen
sub-lattice in $\mathrm{UO_2}$\cite{Faber1976}, and eventually it was
demonstrated that the magnetic structure is 3-$\vec{k}$ type, i.e. Fourier components of all three members of the star of $\vec{k}=\langle001\rangle$ are \emph{simultaneously} present on each magnetic site of the lattice.\cite{Rossat1986}
These Fourier components of the magnetic moment, $\vec{\mu}_k$,
are perpendicular to the
propagation vector $\vec{k}$, so that the structure is \emph{transverse}.
The overall symmetry remains cubic, and the magnetic moments
$\vec{\mu}_{0}$ point
along the $[111]$ directions of the cubic unit cell. As $\vec{\mu}_k$
can assume two mutually orthogonal directions for each $\vec{k}$ value
(for instance, $\vec{\mu}_k$ parallel to $[010]$ or $[001]$ for $\vec{k}$
= $[100]$), there two distinct  
3-$\vec{k}$  transverse
magnetic structures (Figure 1).


\begin{figure}[t!]
\centering
\includegraphics[height=0.7\textheight]{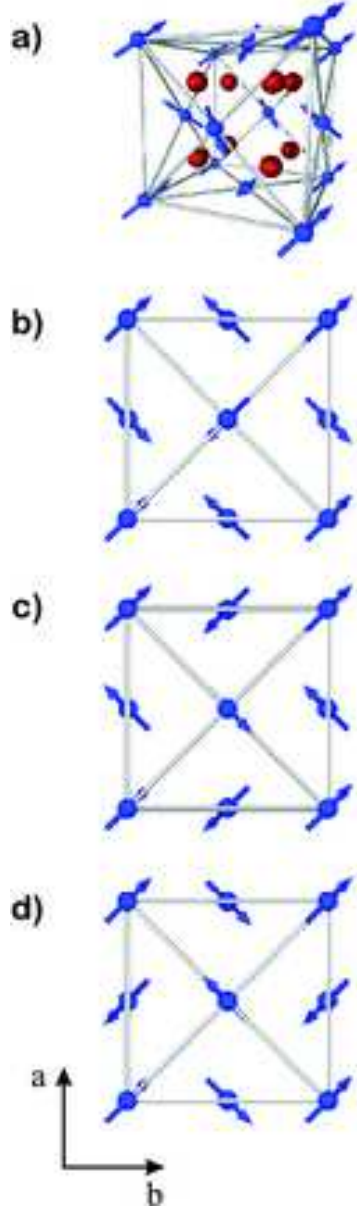}
\caption{
a) Shows a 3d representation of the 3-$\vec{k}$ magnetic structure. Uranium atoms are shown as blue spheres with the corresponding dipole magnetic moment direction shown as an arrow. Oxygen atoms are shown as red spheres. 
b) The longitudional 3-$\vec{k}$ structure as projected onto the $a-b$ plane, i.e. on moving from an atom at (000) with a moment along [111] to the atom at $(\frac{1}{2}\frac{1}{2}0)$ the moment direction changes to $[\overline{1}\overline{1}1]$.
c) and d) are transverse structures. Following the reasoning above the atom situated at $(\frac{1}{2}\frac{1}{2}0)$ has moment directions $[\overline{1}1\overline{1}]$ and $[1\overline{1}\overline{1}]$ for c) and d) respectively.}
\label{Fig1}
\end{figure}

Experiments on powder samples with
chopper spectrometers at spallation neutron sources allowed a precise
and reliable determination of the CF Hamiltonian acting
on the U ions \cite{Amoretti89}, and
experiments with triple-axis spectrometers and axial polarization
analysis have
been performed to study collective excitations in the ordered phase,
and dispersive magnetic excitons in the paramagnetic 
phase\cite{Caciuffo99}.
The experimental picture is therefore exhaustive, but a satisfying 
theoretical
description of low-temperature collective excitations is still
missing.

Magnetic order was expected to occur also in
$\mathrm{NpO_2}$, for which a large
$\lambda$-like anomaly in the
heat capacity\cite{Osborne} and a maximum in the magnetic 
susceptibility
curve\cite{Ross} were observed at $T_{0}=25$~K.
However, neither M\"{o}ssbauer
spectroscopy\cite{Dunlap} nor neutron diffraction
experiments\cite{Cox,Heaton,Boeuf83} have found evidence of
magnetic ordering in this compound. The experiments set an upper limit
on the Np ordered magnetic moment of $0.01\;\mu_{B}$, a value that is
surprisingly small given the Kramers nature of
$\mathrm{Np^{4+}}$ ions ($5f^{3}$, $^{4}I_{9/2}$ in Russell-Saunders
approximation),
and the large effective Curie-Weiss paramagnetic moment
deduced from susceptibility measurements ($\sim 3\;\mu_{B}$).

Indeed, Resonant X-ray Scattering (RXS)
experiments\cite{Paixao02} at
the Np $M_{4}$  edge exclude usual magnetic
dipole ordering at $T_{0}$, but provide direct evidence of long-range 
order
of an electric quadrupole moment with $\Gamma_{5}$ symmetry.
The phase transition is purely electronic and involves neither 
internal
nor external crystallographic distortions,
so the symmetry of the system remains 
cubic\cite{Boeuf83,Caciuffo87,Mannix99}.
It has been suggested that the primary order
parameter is associated with $\Gamma_{4}^{t}$ magnetic octupoles,
ordering in a 3-$\vec{k}$ \emph{longitudinal} structure with
$\langle 001 \rangle$ propagation vector. The ground state
(GS) is a singlet with no static dipole magnetic moment in which 
magnetic
octupole order breaks time reversal invariance and induces an electric
quadrupole as the secondary order parameter\cite{Santini00,Fazekas}.

The magnetic behavior of $\mathrm{U_{1-x}Np_{x}O_{2}}$ solid solutions has
also been extensively
investigated by powder neutron diffraction, $\mathrm{^{237}Np}$ M\"{o}ssbauer
spectroscopy, and magnetic susceptibility
measurements\cite{Caciuffo-EL,Caciuffo-JPL,Tabuteau-SSC}.
The $\mathrm{^{237}Np}$ M\"{o}ssbauer
isomer shift indicates that Np ions are tetravalent,
with $5f^{3}$, $^{4}I_{9/2}$ electronic configuration, independently
of temperature and composition.  Magnetic
order has been detected in the
composition range from $x= 0.15$ and $x= 0.75$, with transition
temperatures decreasing with increasing $x$, from $T_{N} \approx 17$~K
($x$ = 0.25) to $T_{N} \approx 11$~K ($x = 0.75$), showing that
substitution of Np for U reduces the
strength of U-U exchange interactions. For $x= 0.25$, the magnetic
structure is transverse antiferromagnetic type I, as in $\mathrm{UO_{2}}$, with
propagation vector $\langle001\rangle$, and an average ordered magnetic moment
$\mu = 1.8(1) \mu_{B}$. As M\"{o}ssbauer
data are compatible with an ordered magnetic moment
$\mu_{Np} = 0.6(2) \mu_{B}$ on the Np ions, this gives
$\mu_{U} = 2.2(2) \mu_{B}$ on the U ions. For larger Np content, the
magnetic order is short range and incommensurate\cite{Caciuffo-EL}.

\section{Resonant X-ray Scattering}

Resonant X-ray Scattering (RXS) occurs when a  
photon is
virtually absorbed by
exciting a core electron to empty states, and subsequently re-emitted
when the excited electron and the core hole
recombine\cite{Hannon,Hill}. This process
introduces anisotropic contributions to the X-ray susceptibility
tensor\cite{Blume2}, the amplitude of which increases dramatically as
the photon
energy is tuned to an atomic absorption edge. In the presence of 
long-range
order of magnetic moments, electronic orbitals occupancy, or spatially
anisotropic valence-electron clouds, the interference of the anomalous
anisotropic scattering amplitudes may lead to the excitation of Bragg
peaks at positions forbidden by the crystallographic space group.
For absorbing atoms
belonging to the  actinide series, both L- and M-edges
are of practical use, as their energy is large enough
for Bragg diffraction to be allowed in most actinide compounds. The
experiments reported here have been performed at the U and Np $M_{4}$ and
$M_{5}$ edges, involving respectively $3d_{3/2} 
\leftrightarrow 5f$
and
$3d_{5/2} \leftrightarrow 5f$ electric-dipole, E1, virtual excitations.

The amplitude of \emph{magnetic} scattering from a given atom at the
$3d-5f$ threshold is the product of a geometrical factor, depending 
on the
polarization of the incident ($\vec{\epsilon}$) and scattered
($\vec{\epsilon}^\prime$)
photons, and a linear combination of resonance strength
factors,$F_{L,M}$,

\begin{equation}
F^{[1]} \propto (\vec{\epsilon} \times \vec{\epsilon}^\prime) 
[F_{1,1}-F_{1,-1}]
	\label{Eq1}
\end{equation}

Magnetic scattering therefore requires a rotation of the photon
polarization, and a difference in the transition probabilities to
intermediate states with $M=\pm \;1$. This asymmetry can arise because
of the net spin polarization of the $5f$ states, or from a difference 
between
overlap integrals, resonant energy, or lifetime for spin-up and 
spin-down
channels\cite{Hill,Blume,Gibbs}.

RXS can also signal long-range order of anisotropic charge
distributions. It is the asphericity of the
atomic electron density that generates the anomalous tensor component
in the atomic scattering factor. If the electron clouds on different
sublattices have different orientations, superlattice reflections
occur due to the reduced translational symmetry. This is the
so-called Templeton or ATS (Anisotropic Tensor Susceptibility) scattering \cite{Templeton85}
and is observed, for
instance, in case of antiferro-electric order of the quadrupole 
moments.
The scattering amplitude arising from E1 transitions can be
described by second-rank tensors $f(\vec{r})$, invariant under the 
point symmetry
of the scattering atom.  If $\tilde{f}(\vec{Q})$ is the Fourier
transform of $f(\vec{r})$, the scattering amplitude can be written as
\begin{equation}
F^{[2]}(\vec{Q}) \propto \vec{\epsilon}^{\;\prime}
\cdot \tilde{f}(\vec{Q}) \cdot \vec{\epsilon},
	\label{Eq2}
\end{equation}
where $\vec{\epsilon}$ and $\vec{\epsilon}^{\;\prime}$ are the 
polarization
vectors of the incident and scattered beam, where the rotation about 
the azimuth is implicitly included in the electric field vector.

\section{Experimental details}

$\mathrm{U_{0.75}Np_{0.25}O_{2}}$ single crystals have been grown using the
chemical-transport reaction method at the Institute for Transuranium 
Elements
in Karlsruhe. Starting materials were obtained by stoichiometric
co-precipitation by
ammoniac, calcination under air at 1073~K, followed by reduction under
Ar/H$_{2}$ ($5\%$) at 1073~K. Pellets were fabricated and 
sintered
at 1973~K for 6 hours in Ar/H$_{2}$ atmosphere.
Charges of about 5~g of starting material were crushed and welded
in a quartz tube with TeCl$_{4}$ (4.5~mg/cm$^{3}$) as
transport agent in a $10^{-6}$ torr vacuum. The feed substances were 
maintained
at 1323~K and transported under the influence of the temperature 
gradient
into the growth region at 1223~K for 14 days.
The single crystals obtained were characterized by chemical analysis,
scanning electron microscopy,
single crystal x-ray diffraction and x-ray diffraction on a powder
obtained from crushed crystals. No differences in the average 
composition or evidence for clustering effects have been observed by 
these techniques. However, differences in the spatial homogeneity of the Np ions 
distribution in individual crystals grown by the method described 
above cannot be excluded. 

Two specimens with a mass of 5.20~mg and 3.63~mg have been selected to measure
the heat capacity as a function of temperature and magnetic field
amplitude, using a PPMS-9 Quantum Design platform.
The RXS experiment has been performed
on the ID20 beamline
of the European Synchrotron Radiation Facility (ESRF) in Grenoble,
France. For these measurements we used a crystal 0.204~$\mathrm{mm}^{3}$ in volume, with a mass of 2.26~mg and the $[111]$ axis perpendicular
to the exposed face. The sample was of excellent quality with a
mosaic width of $0.05~\deg$. For safety requirements, the crystal was
encapsulated in a sealed double-wall container with beryllium windows.

The ID20 spectrometer exploits the fundamental harmonic emission of
two 42-mm-period undulators. A double crystal Si(111) curved
monochromator provides sagittal focusing and selects the appropriate
X-ray energy.
Vertical focusing and filtering of higher harmonics
is achieved by Si mirrors,
that deliver at the sample position a 99 $\%$ linearly
polarized beam, with
polarisation perpendicular to the scattering
plane ($\sigma$) and incident wavevector $\textbf{k}_{i}$.
A mosaic crystal positioned between
sample and detector was used to analyze
whether the polarization of the scattered beam (with wavevector
$\textbf{k}_{f}$) is parallel
($\pi$) or perpendicular ($\sigma$) to the
vertical scattering plane.

A closed-cycle refrigerator equipped with an azimuthal-rotation stage
provided a base temperature of 10~K and the possibility to perform
azimuthal scans.
The spectrometer was operated in the photon energy interval from 3.5
to 4.0~keV, including the U and Np $M$-edges
(U: $E_{M_{5}} = 3.552$~keV, $E_{M_{4}} = 3.728$~keV; Np: $E_{M_{5}} = 3.664$~keV, $E_{M_{4}} = 3.845$~keV), using
the $(1 1 1)$ reflection from a Au analyzer (Bragg angle
of $43.2\deg$ at the Np $M_{4}$ edge,
and $41.6\deg$ at the U $M_{4}$ edge; no correction was performed to
account for the departure from the ideal $45\deg$ orientation
of the analyzer crystal). The energy resolution
was about 0.5~eV (full width at half maximum), and the beam flux
at 3.845~keV was estimated as $5 \times 10^{11} \mathrm{photons\;s^{-1}}$.
Integrated intensities $I_{m}$ for Bragg reflections were obtained
numerically, by fitting a
linear background and a Lorentzian squared lineshape to
the measured peaks. Corrections for the size of
the footprint of the incident beam
and for absorption were applied as
$I_{c} = I_{m}~\mu (1 + \sin \alpha/\sin \beta)$,
where $\alpha$ and $\beta$ are the angles formed by the incident and
scattered beam directions with the sample
surface, and $\mu$ is the energy dependent
absorption coefficient. Because of the strong absorption all
reflections must be found in Bragg geometry.

\section{Results of the Experiment}

Figure 2 shows the 
heat capacity as a function of temperature measured for the
$\mathrm{U_{0.75}Np_{0.25}O_{2}}$ sample 
with 5.20~mg mass.
Data were collected from 0.7 to 300~K in zero field and under an 
external
magnetic field of 9~T.
The base temperature was limited by self-heating effects,
that become sizeable below 3~K. Contributions to the measured heat 
capacity
from the sample platform (including grease used to establish thermal
contacts) were measured separately and subtracted from the total 
signal.
The results reveal the presence of a $\lambda$-like cusp at 18.9~K, 
and
two small anomalies at 16.8~K, and 15~K.
The application of a 9~T magnetic
field does not produce appreciable effects; the maximum at 18~K
undergoes a very small shift to a lower temperature while its 
amplitude
slightly decreases. The
specific heat of the isostructural compound $\mathrm{ThO_{2}}$ is assumed to 
be a
close estimate of the vibrational contribution, and is shown as a 
solid
line in Figure 2. As previously reported for both
$\mathrm{UO_{2}}$ and $\mathrm{NpO_{2}}$\cite{Osborne},
the magnetic contribution is appreciably different
from zero at temperatures far above the anomaly also in the solid solution.

The broken line in Figure 2 represents the results of measurements 
performed on a second single-crystal with the same nominal composition 
of the first one.
The curves obtained for the two samples are quite different; for the 
second one, the main anomaly occurs at 16.8~K, whereas small 
anomalies are visible at 18.4~K and 15~K.
The characterization analysis suggests the same stoichiometry within 
experimental errors, and X-ray diffraction does not show line 
broadening, thus excluding major clustering effects. On the other hand, the 
subtle interplay between magnetic and electric multipolar interactions 
in this system could be very sensitive to even small changes in the 
chemical short-range order. The large 
difference observed in the heat capacity of the two samples could 
therefore reflect small variations in the probability distribution 
of Np neighbors around U ions. 

\begin{figure}
\includegraphics[width=\columnwidth]{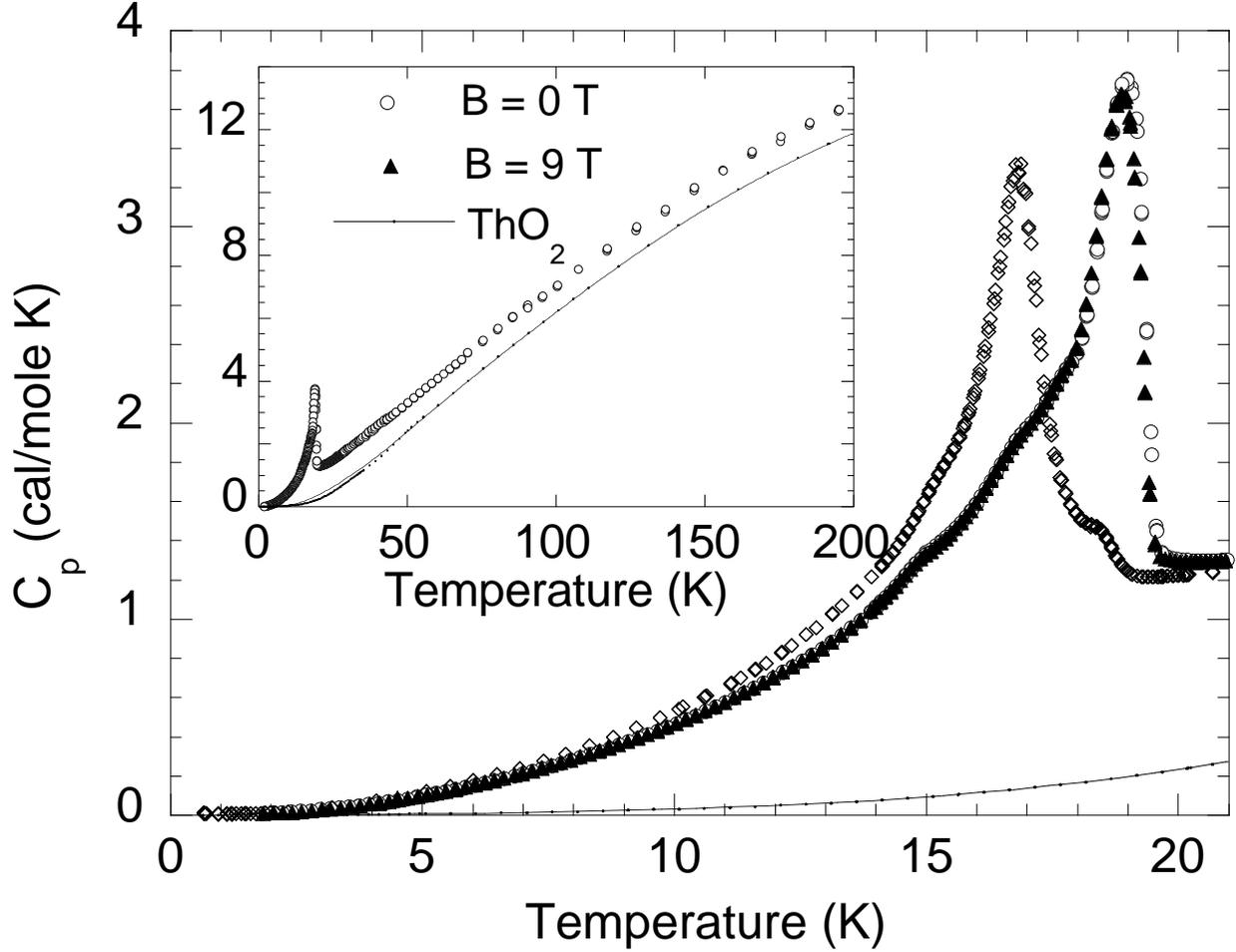}
\caption{Heat capacity measurements performed with a Quantum Design
PPMS-9 system. Data were taken in an applied field of (circles)
0~T and (triangles) 9~T.
A $\lambda$-like peak and two anomalies are clearly visible. The 
solid line
is a fit to heat capacity data measured for the non-magnetic
iso-structural compound $\mathrm{ThO_{2}}$. The broken line corresponds to 
results obtained for a second sample with the same nominal composition.}
\label{fig:cp}
\end{figure}

The presence in the diffraction pattern of resonant superlattice
reflections
with propagation vector $\langle 001 \rangle$, forbidden in the
paramagnetic phase, proves that long-range order of some multipolar
moment characterizes the low-temperature ground state.
The integrated intensity of the $(112)$ Bragg peak at 10~K
is shown as an example in Figure 3, as a function of the photon
energy across the U and the Np $M_{5}$ and $M_{4}$ absorption
edges. These data were collected
without analyzing the polarization of the diffracted beam, and
therefore represent the sum of
the intensities in the \sigmasigma\ and \sigmapi\ polarization channels. The spectrum can
be reproduced as the superposition of four oscillators that
interfere with each other. The solid line in Figure 3 is obtained
by assigning to each oscillator a
Lorentzian lineshape centered at the E1 electric-dipole energy,
including corrections for self-absorption and
energy dependence of the beam attenuation.

\begin{figure}
\includegraphics[width=\columnwidth]{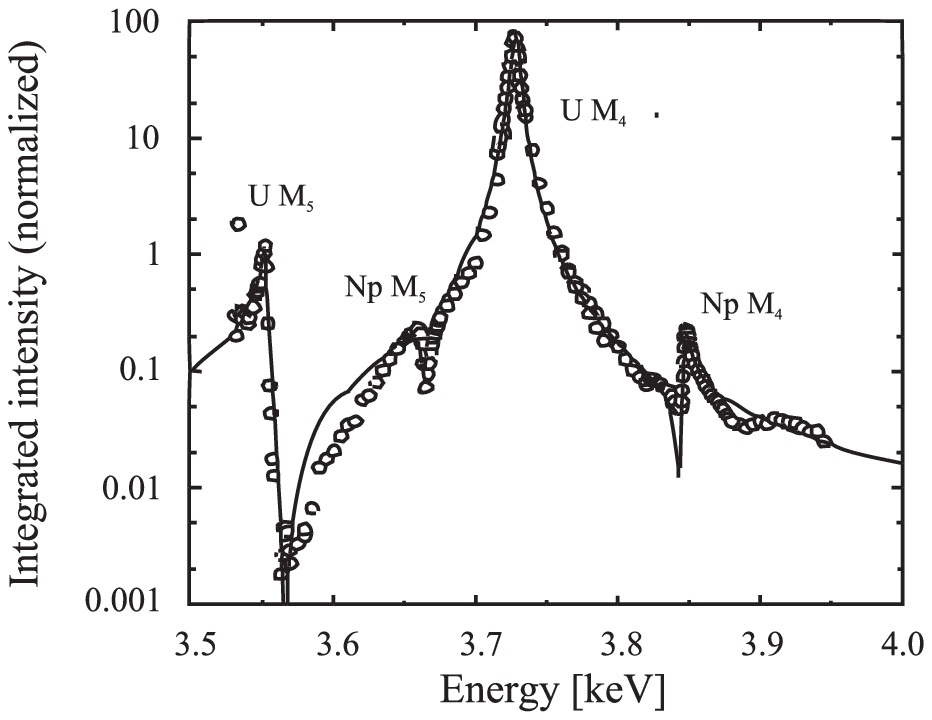}
\caption{Normalized integrated intensity as a function of energy for
the (112) reflection of \unpo at $10\;K$.
Polarization analysis was not performed.
The solid line is a simulation assuming four dipole oscillators,
each represented by a Lorentzian function.}
\label{fig:all-res-nopa}
\end{figure}

At 10~K, a resonant behavior has been observed for both the
\sigmapi\ and
the \sigmasigma\ components of the spectrum.
This is shown in Figures 4 and 5,
around the U and the Np $M_{4}$ edge, respectively. Note that the \sigmasigma\ have Lorentzian squared lineshapes, and are shifted by $-2$~eV wrt. the \sigmapi\ peak. This has been previously observed in systems where the \sigmasigma\ signal is known to originate from the $F^{[2]}$ term.\cite{PRB} 
The ratio between the intensities of the
\sigmasigma\ and the \sigmapi\ peak is about 30\% for Np,
and about 3\% for U. For strictly magnetic \textit{dipole}
ordering, there is no signal in the \sigmasigma\ channel. Thus, the
finite ratio indicates that the
structurally forbidden peaks do \emph{not} stem \emph{uniquely} from
dipole magnetic-order. Our RXS results are
compatible with dipole magnetic-order together with
an ordered distribution of electric quadrupole moments. Resonant
scattering from orientationally ordered aspherical electron clouds
appears in both polarization channels. On the other hand,
the shape of the resonance across the $M_{4}$ absorption edge is
different for the \sigmasigma\ and the \sigmapi\ contributions.
Whilst the \sigmapi\ peak
can be fitted to a Lorentzian function, the \sigmasigma\ resonance has
a Lorentzian-squared lineshape. A Lorentzian peak is
usually observed when the intensity enhancement is due to dipole 
magnetic
order, whereas a Lorenztian-squared peak has been observed in
$\mathrm{NpO_{2}}$, both in the rotated and the unrotated polarization channel\cite{Paixao02}.

The narrow energy profile of the Bragg intensity at the Np $M_{4}$ edge in $\mathrm{NpO_{2}}$ has been associated with a $^{3}d _{3/2}$ core-level structure, and attributed to magnetic and charge contributions to the E1 Bragg amplitude described by Np 5f multipoles of rank 3 and 4 \cite{SWL_NpO2}. This explanation cannot be applied in the present case, as $\mathrm{U^{4+}}$ ions in $\mathrm{UO_{2}}$ do not carry magnetic octupoles. On the other hand, in ref. \onlinecite{Paixao02} the RXS signal observed in $\mathrm{NpO_{2}}$ is interpreted entirely as a consequence of electric-quadrupole order, both for the \sigmasigma\ and the \sigmapi\ channels, as the magnetic ordered moment in $\mathrm{NpO_{2}}$ is vanishingly small. The results presented here seems to support these latter interpretations.

As the magnetic ordered moment in $\mathrm{NpO_{2}}$ is vanishingly small, in
that case the resonant scattering is entirely due to
electric-quadrupole order. It therefore appears that the Lorentzian
squared shape is a characteristics of quadrupole order in these
compounds. We conclude that in this solid solution, the \sigmapi\ signal is
dominated by the magnetic dipole contribution, whilst the \sigmasigma\ resonance is due only to the electric-quadrupole order.

\begin{figure}
\includegraphics[width=0.75\columnwidth]{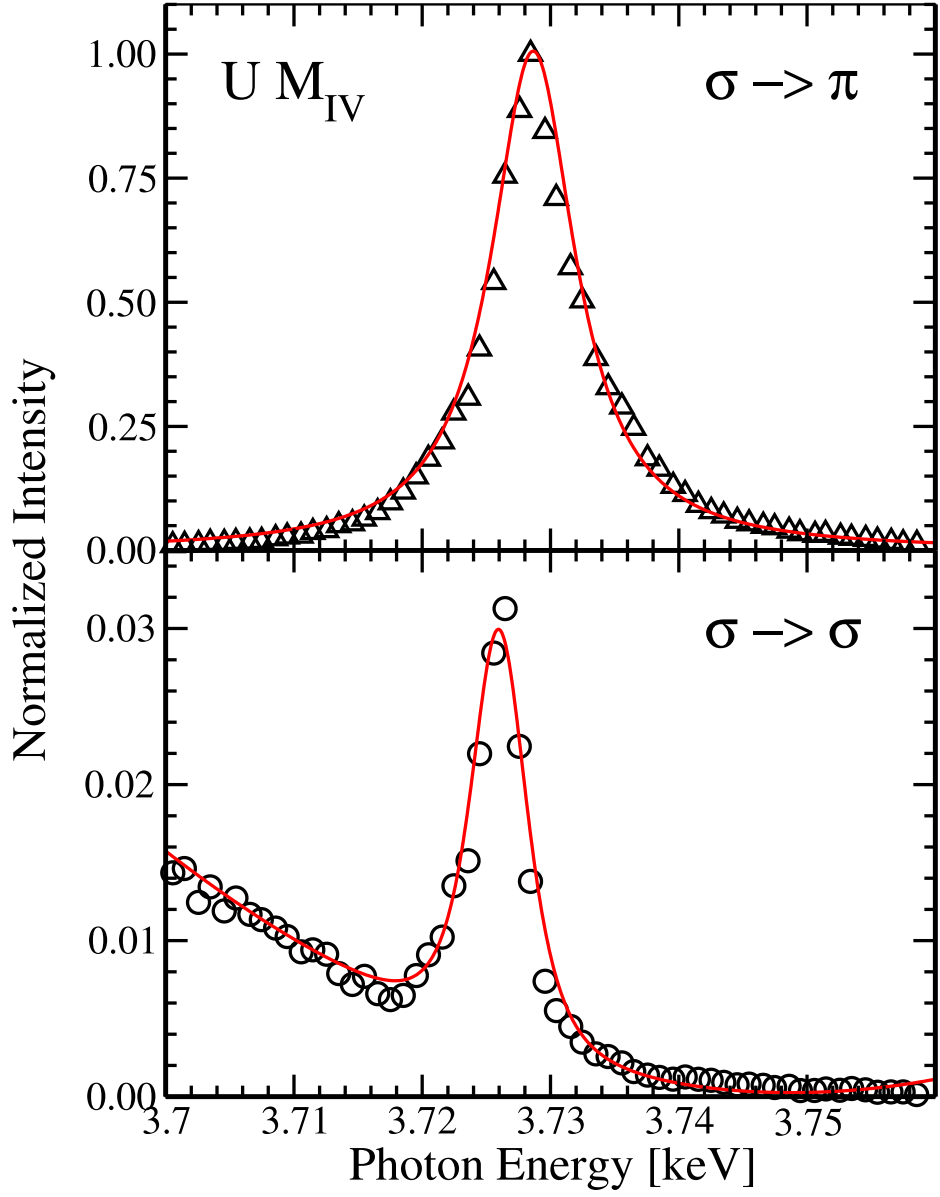}
\caption{Integrated intensity of the (112) superlattice reflection,
at 10~K, as a function of photon energy in the vicinity of the U
M$_{4}$ absorption edge. Upper panel \sigmapi\,
lower panel \sigmasigma. The solid line (upper panel) is a
fit to a Lorentzian lineshape with a FWHM of 8~eV.  In the lower panel the solid line represents a fit to a Lorentzian squared lineshape with the same width and a sloping background.}
\label{fig:U-res}
\end{figure}

\begin{figure}
\includegraphics[width=0.75\columnwidth]{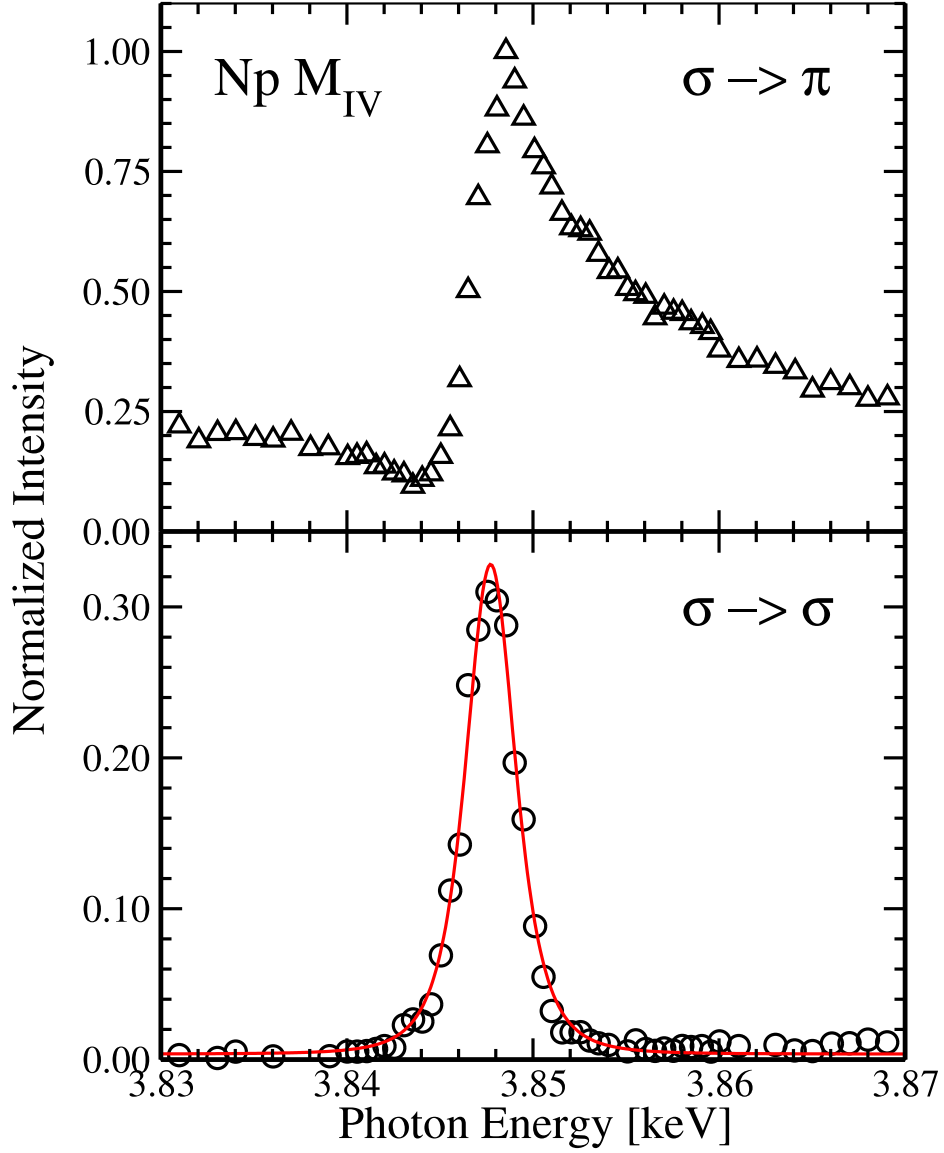}
\caption{Integrated intensity of the (112) superlattice reflection, 
at 10~K,
as a function of photon energy in the vicinity of the Np M$_{4}$
absorption edge. Upper panel \sigmapi\, lower panel \sigmasigma.
The solid line (lower panel) show a fit to a Lorentzian squared 
lineshape.
The \sigmapi\ contribution is dominated by the magnetic dipole signal
and exhibits a Lorentzian lineshape (upper panel). Intensities
are normalized to the maximum of the \sigmapi\ signal at the M$_{4}$
dipole threshold. Note that the \sigmapi\ signal (upper pannel) is strongly influenced
by the dominant signal from the U M$_{4}$ resonance,
see Figure \label{fig:all-res-nopa}}
\label{fig:Np-res}
\end{figure}

The results of $\Psi$ (azimuthal) scans about the $(112)$
magnetic peak, taken at $T=10$~K with photons energy tuned to the Np
and U $M_{4}$ edges, are shown in Figure 6.
The intensity of the peak is
measured while the sample is rotated about the scattering vector
$\vec{Q}$, kept constant at the chosen value. The origin of 
the
azimuthal angle, $\Psi$, is defined where the 
$[11\overline{1}]$-direction is within the
scattering plane. Oscillations with two-fold symmetry are observed for both the U and Np signals in the \sigmasigma\ channel. On the
other hand, the \sigmapi\ intensities are independent of $\Psi$. The azimuthal dependence of the peak
intensity is similar at the U and Np $M_{4}$ edges,
indicating the same electric quadrupole orientation on the two ions.

\begin{figure}
\includegraphics[width=\columnwidth]{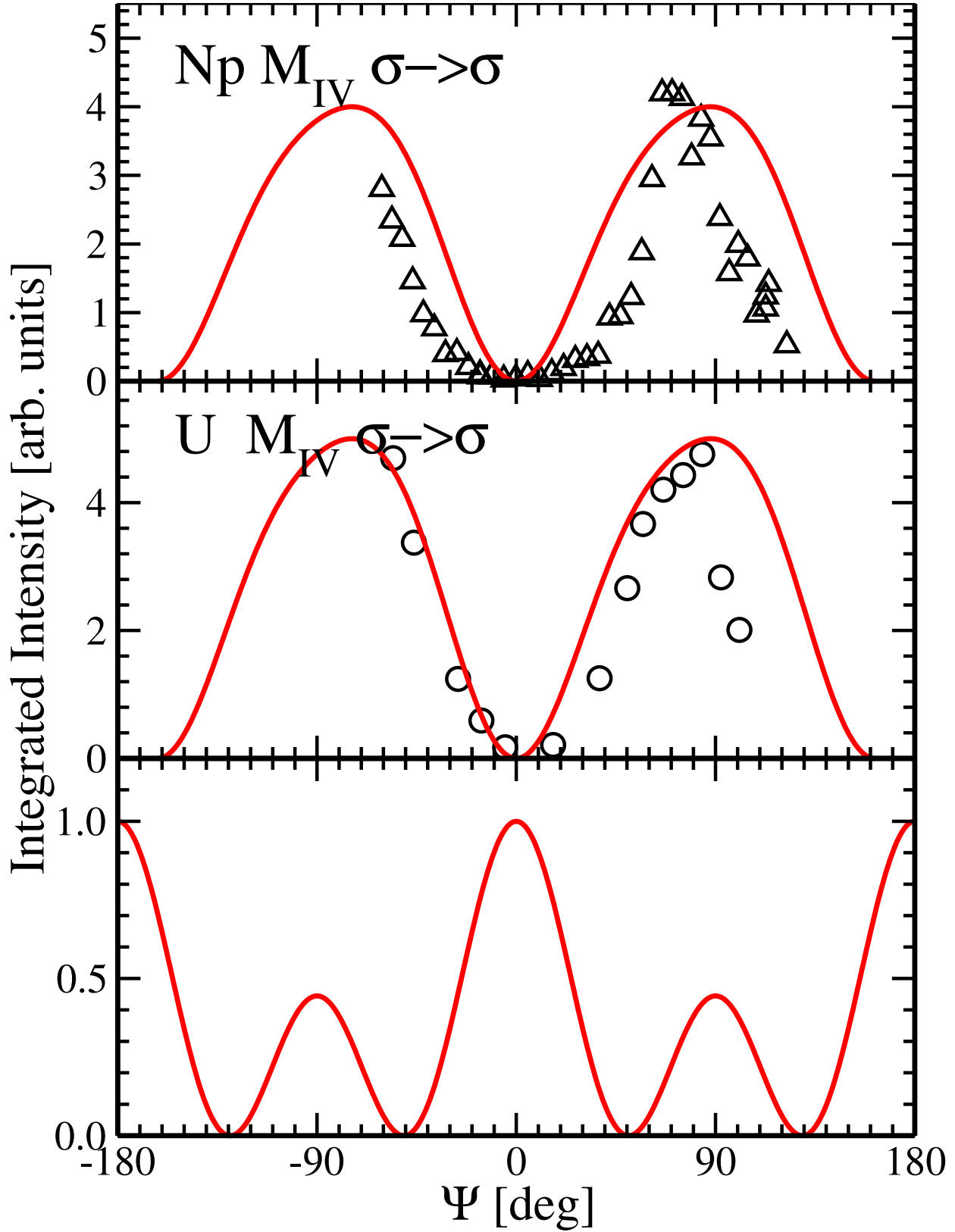}
\caption{Azimuthal dependence of the integrated intensity of the (112)
superlattice reflection in the \sigmasigma\ channel at both the Np 
M$_{4}$
(top panel) and U M$_{4}$ (middle panel) absorption edges.
Lines are intensities calculated assuming a electric quadrupole 
moments along
the $\langle111\rangle$ direction, for a addition of two noncollinear transverse antiferro-quadrupolar domains (see Figure~\ref{Fig1}). The lower panel shows a simulation for a longitudinal 3-$\vec{k}$ structure for comparison.}
\label{fig:azimuth}
\end{figure}

The solid line in Figure~\ref{fig:azimuth} is a simulation of the 
scattered intensity for a 3-$\vec{k}$ anti-ferroquadrupole 
structure assuming a incoherent addition of the two transverse s-domains (See Figure~\ref{Fig1}). The lower panel of Figure~\ref{fig:azimuth} shows a simulation of the longitudional 3-$\vec{k}$ anti-ferroquadrupole structure. This model clearly does not fit the data. Details of the calculations used to obtain the simulated azimuthal dependence are given in the appendix.

Figure 7 shows the temperature
dependence of the $(112)$ integrated intensities,
normalised to unity at 10~K. The reported data were taken at the
M$_{4}$ dipole threshold of either Np or U, for both polarization
channels. Intensity was collected as a function of time while the
sample temperature was varied at a rate of 2~$\mathrm{K\;min^{-1}}$.
The \sigmapi\ intensity is
dominated by the magnetic dipole contribution and is therefore 
proportional
to the square of the ordered magnetic moment.
The \sigmasigma\ channel intensity, on the other hand, is proportional
to the square of the electric quadrupole order parameter.

On cooling, the magnetic-dipole order develops first on U at
$T_{N}\approx 19$~K; the 
order-parameter increases with an apparent 
plateau between $T_{N}$ and 18~K. 
The magnetic order on Np, and electric quadrupole order of both ions,
is established at $T_{N}$ but the corresponding order-parameters 
increase at a lower rate. 
Note
that the AF phase transition
in $\mathrm{UO_{2}}$ is strongly discontinuous, whereas in the solid solution the
quadrupolar and the magnetic phase transition are continuous.
The temperature dependence of the U \sigmapi\ and \sigmasigma\ 
intensity is
remarkably different, whilst for Np the corresponding difference is marginal. 


\begin{figure}
\includegraphics[width=\columnwidth]{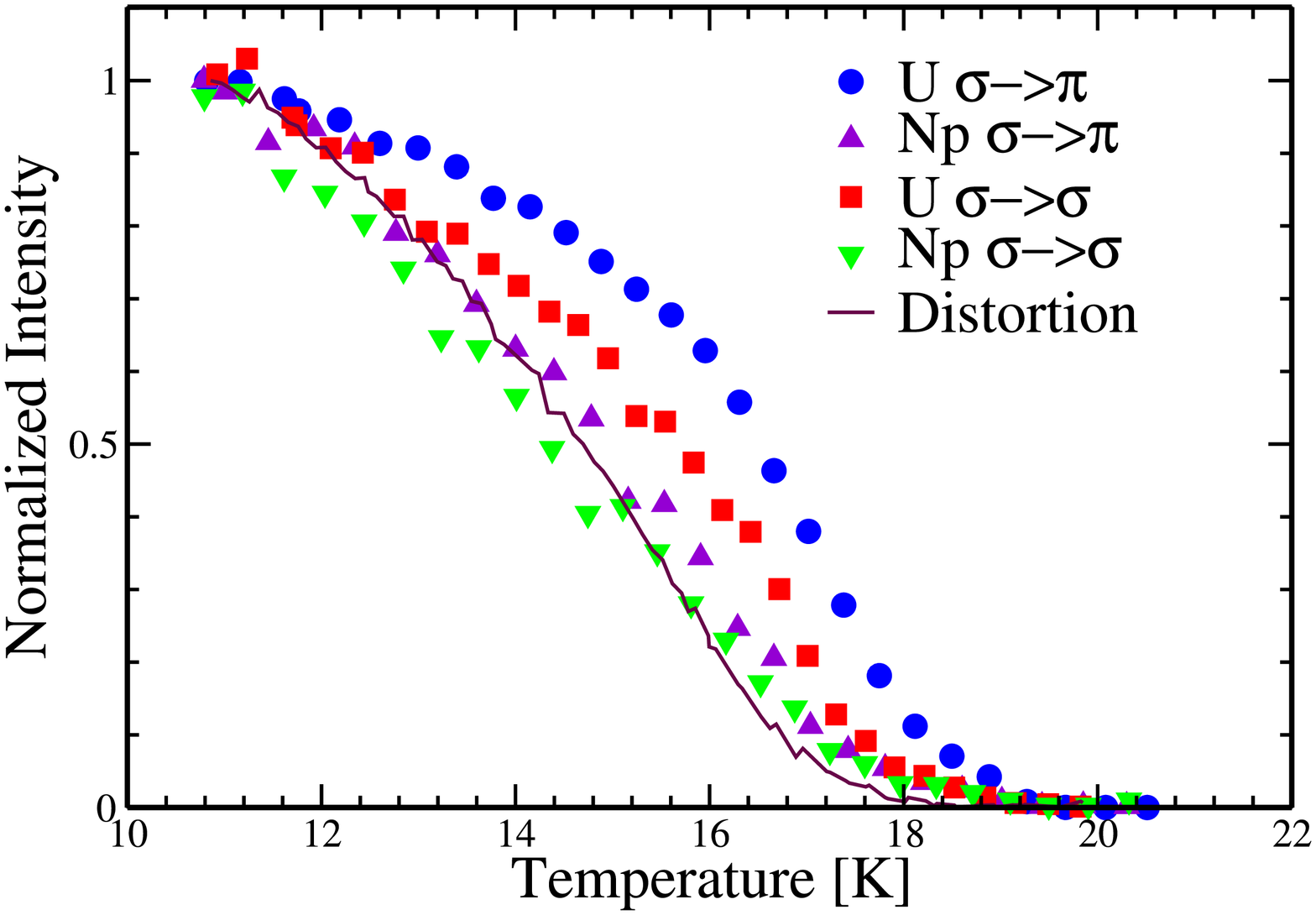}
\caption{Scattered intensity of the forbidden reflections as a 
function
of temperature. The data for the (112) reflection were taken at the M$_{4}$ edges of U and Np in both polarization channels: (filled circles)
U \sigmapi\ intensity; (open circles) U \sigmasigma\ intensity;
(upper triangles) Np \sigmapi\ intensity; (down triangles) Np 
\sigmasigma\
intensity. The solid line intensities are measured
for the $(421)$ internal lattice modulation at 7.5~keV, indicating the
onset of a Jahn-Teller distortion
at $T_{N}$. Solid lines are fit to a power law.}
\label{fig:dspacing}
\end{figure}

Besides resonant superlattice reflections, diffraction peaks 
corresponding
to a deformation of the
oxygen cube have also been found. The angular positions of the
observed peaks corresponds to those expected for the
inhomogeneous internal distortion observed in
$\mathrm{UO_2}$\cite{Faber1976}, and resulting from combinations of
normal modes of the oxygen sub-lattice.
For these measurements, we used a photon energy of
7.5~keV
and a Ge (333) analyser to select the \sigmasigma\
polarisation channel, where structural Bragg peaks give a 
contribution.
As an example, we show in Figure 7 (solid line) the
intensities measured
for the internal lattice modulation measured at the (421) reflection. It appears that
the onset of the Jahn-Teller distortion
coincides with the onset of the \sigmasigma\ scattered intensity, i.e. the long range order of the electric quadrupolar moments.

A departure from the
cubic symmetry was also checked by following
angular position and width of the $(006)$ lattice reflection, but no evidence for a lowering of symmetry was found. This confirms the 3-$\vec{k}$ nature of the ordering. On the
other hand, as shown in Figure 8,
an anomalous volume contraction of the cubic cell is observed on
cooling between $T_{N}$ and 18~K.

\begin{figure}
\includegraphics[width=\columnwidth]{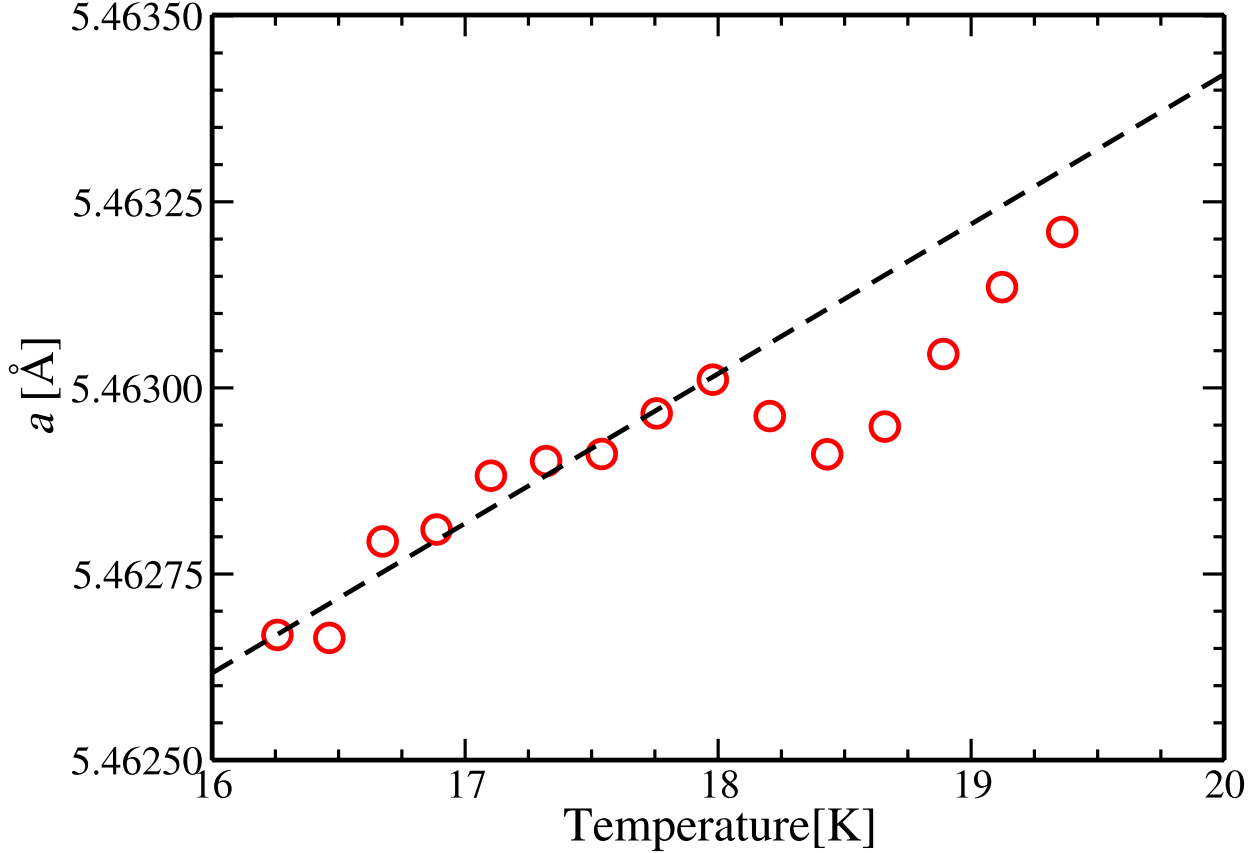}
\caption{Temperature dependence of the lattice parameter of the
$U_{0.75}Np_{0.25}O_{2}$ solid solution.
Error bars are smaller than the symbols. A anomalous contraction can 
be
observed in the temperature interval between $18$ and $20\;K$.}
\label{fig:dspacing}
\end{figure}

\section{Discussion}

The CF spectrum of actinide dioxides is known with high degree of
precision from inelastic neutron scattering (INS) experiments. The
CF ground state of U in $\mathrm{UO_2}$ is a $\Gamma_5$ triplet, whereas the
CF ground state of Np in $\mathrm{NpO_2}$ is a $\Gamma_8$ quartet. The CF
potential changes little among $\mathrm{UO_2}$, $\mathrm{NpO_2}$, and even $\mathrm{PuO_2}$.
Therefore, the CF acting upon the U and Np ions in solid solutions
$\mathrm{U_xNp_{1-x}O_2}$ is expected to be the same as in the pure
compounds.

The $\Gamma_5$ triplet carries magnetic-dipole and
electric-quadrupole degrees of freedom. U-U dipolar and
quadrupolar interactions are provided by superexchange (SE) and
additional U-U quadrupolar interactions are carried by phonons
through the so-called virtual-phonon exchange. In $\mathrm{UO_2}$ dipolar
interactions are stronger than quadrupolar interactions and would
produce on their own a second-order AF transition at around 25 K.
A quadrupole moment is induced by dipoles as secondary order
parameter (OP). In turn, this quadrupole induces a lattice
distortion of the oxygen cage. Although quadrupolar interactions
are too weak to produce a purely quadrupolar phase transition,
they contribute to the free energy of the system when quadrupoles
are induced by dipoles. This contribution is believed to be
responsible for the change of the magnetic transition from second-
to first-order with an increase of $T_N$ up to about 31 K. The AF
structure, type-I transverse 3-$\vec{k}$, is probably
stabilized by quadrupolar interactions. As yet no theoretical model
has been able to explain the occurrence of the 3-$\vec{k}$
structure.

In $\mathrm{NpO_2}$, the $\Gamma_8$ quartet carries magnetic-dipole,
electric-quadrupole and magnetic-octupole degrees of freedom.
Np-Np dipolar, quadrupolar and octupolar interactions are provided
by superexchange (SE) and additional Np-Np quadrupolar
interactions are carried by phonons. The latter are expected to be
appreciably smaller than the corresponding U-U interactions
because of the smaller quadrupolar polarizability of Np ions.
Dipole-dipole interactions are expected to be smaller as
well\cite{santini02}. Octupolar interactions appear to dominate,
and induce a second-order AF transition with no dipole moment, but
with an associated quadrupole secondary OP. No lattice distortion
occurs\cite{Paixao02}.

Ion-ion interactions in solid solutions $\mathrm{U_xNp_{1-x}O_2}$ are
particularly rich. As a reasonable guess, U-U and Np-Np
interactions may be assumed to be similar to those in pure
compounds. The lack of a scaling framework for SE makes U-Np
interactions difficult to estimate {\it a priori}. A few qualitative
observations are formulated in the following from a
virtual-crystal-approximation-like perspective, i.e. focusing on
''average'' properties and neglecting effects specifically
associated with the disorder in the sample.

Since the magnetic phase transition of $\mathrm{U_{0.75}Np_{0.25}O_2}$
occurs at a lower $T_N$ than in $\mathrm{UO_2}$, Np acts  as a
magnetic diluent, and therefore U-Np dipolar interactions must be
much smaller than the corresponding U-U interactions. Since adding
Np makes the transition second-order, quadrupolar interactions
must be smaller than in $\mathrm{UO_2}$.  U-Np octupolar interactions are
absent since octupoles are quenched in U. Thus, the phase
transition of $\mathrm{U_{0.75}Np_{0.25}O_2}$ is driven mainly by U-U
dipolar interactions. U quadrupoles (and the associated lattice
distortion) follow U dipoles as secondary OPs but their
contribution to the free energy is too small to affect the phase
transition qualitatively (i.e. to change the value of $T_N$ and
the order of the transition). Their slow rate of growth below
$T_N$ reflects a large $\beta$ exponent, which is typical of a
secondary OP linearly coupled to the square of the primary OP (in
mean-field $\beta$ is twice the primary's OP $\beta$).

Np ions are expected to play a more passive role. The leading
interaction of $\mathrm{NpO_2}$, i.e. octupole-octupole, has no significant
effects in $\mathrm{U_{0.75}Np_{0.25}O_2}$ since it involves the
minority Np-Np pairs but not U-Np and U-U pairs. As U ions
order, Np ions feel the developing U dipoles and quadrupoles
through U-Np interactions, and themselves acquire ordered dipoles
and quadrupoles. The delay by which these Np moments are detected
by X-rays can be understood quite naturally within this picture:
since the U-Np dipolar interaction by which the U dipolar order is
transmitted to  Np ions is (relatively) weak, thermal fluctuations
on Np ions are quenched by the phase transition more slowly than
on U ions. So, Np moments appear at $T_N$ together with U moments,
but they grow more slowly as $T$ decreases, although
eventually as $T\rightarrow 0$ the full moments must be recovered.

In order to check this qualitatively, we performed calculations
for the solid solution based on the virtual-crystal-approximation
(VCA)\cite{jensen}. Russell-Saunders coupling was assumed, and for
simplicity only the CF ground multiplets of U and Np were included
in the calculation, neglecting therefore excited CF states.
Although the wavefunctions of the $\Gamma_8$ quartet of Np depends
on the cubic CF parameter $x$, results do not depend qualitatively
on $x$ ($x=-0.75$ was used\cite{Santini00}). The VCA calculation
assumed a 3-$\vec{k}$ structure and contained four order
parameters to be determined self-consistently at each temperature,
i.e. the size of the dipole moment along $\langle111\rangle$ and of the axial
quadrupole moment along $\langle111\rangle$ for U and Np, $M_{111}(U)$,
$M_{111}$(Np), $Q_{111}$(U), $Q_{111}$(Np). There are six parameters,
i.e. the dipole-dipole ($J$) and quadrupole-quadrupole  ($K$)
interaction strengths for U-U pairs, for U-Np pairs and for Np-Np
pairs. Np-Np couplings play a minor role since these pairs are
in minority. The two leading dipole-dipole couplings $J$(U-U) and
$J$(U-Np) are determined by the known transition temperatures of
pure $\mathrm{UO_2}$ and of the solid solution. If quadrupole-quadrupole
interactions are assumed to be mainly lattice-transmitted, then
$K$(U-Np) and $K$(Np-Np) can be obtained by scaling $K$(U-U) by
using the  known quadrupolar polarizability of Np ions. $K$(U-U)
is the parameter most difficult to determine, because the role of
quadrupole interaction in the 3-$\vec{k}$ structure of pure
$\mathrm{UO_2}$ is not quantitatively clear.

\begin{figure}
\includegraphics[width=\columnwidth]{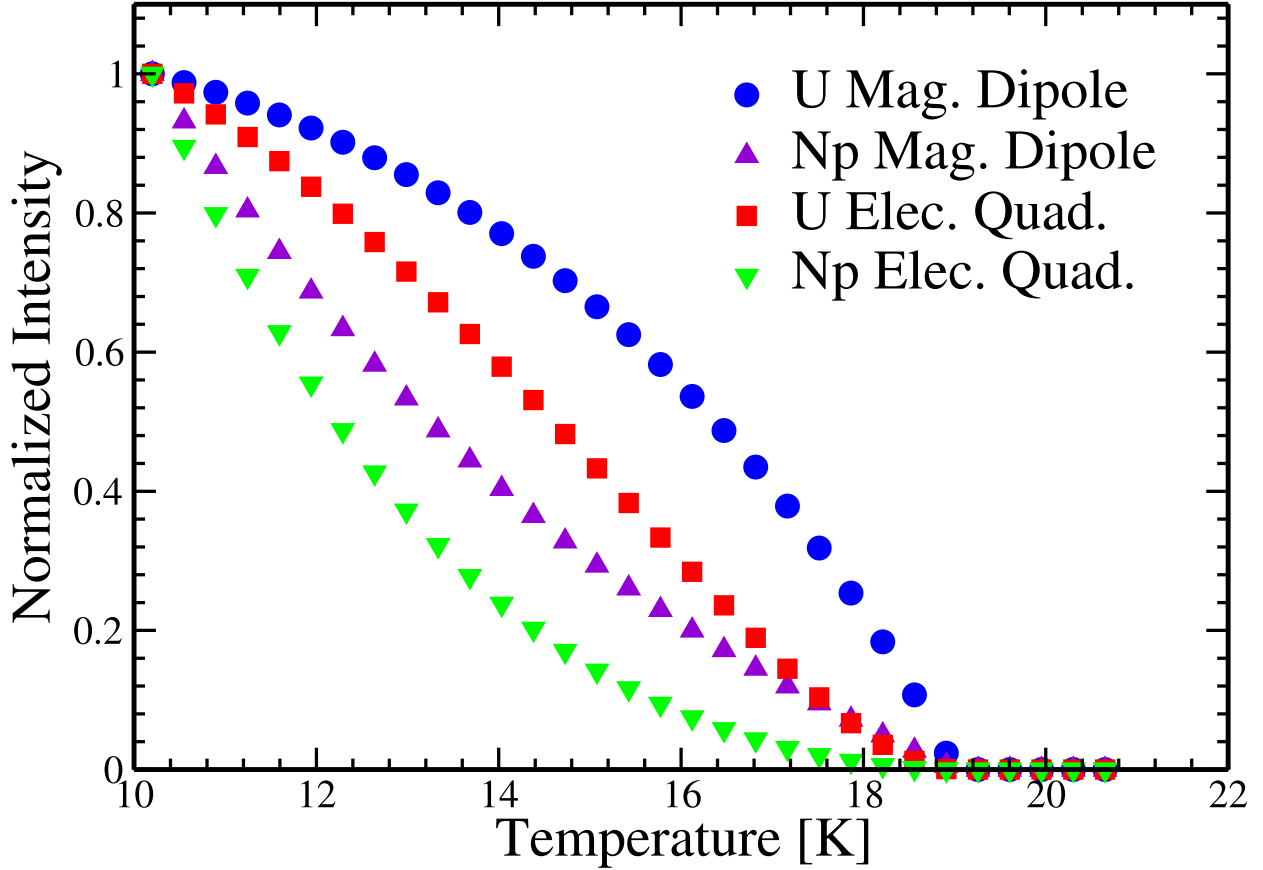}
\caption{Calculated temperature dependence of the normalized RXS
intensity from U and Np dipoles and quadrupoles for
U$_{0.75}$Np$_{0.25}$O$_{2}$.}
\label{fig:vca}
\end{figure}

As an example, we show in Figure~\ref{fig:vca}  calculated RXS intensities for
a parameter set with a fairly small value of $K$(U-U), i.e. about
half the value which had been tentatively used in
$\mathrm{UO_2}$\cite{giannozzi}. $\beta$ exponents are mean-field like in
the VCA, $\beta = 0.5$ for U and Np dipoles and $\beta = 1$ for U
and Np quadrupoles. The curves are in qualitative agreement with
experiment, with U dipoles ordering more quickly than all other
moments. This remains true if $K$(U-U) is varied. Increasing
$K$(U-U) mainly increases the curvature of the red squares in the
figure, i.e. it accelerates the ordering of U quadrupoles.
However, U dipoles always prevail near $T_N$, because of their
smaller $\beta$ exponent.

As far as the specific heat $C$ is concerned, a quantitative
understanding of the measured curve is difficult for at least two
reasons. The first is that the sample dependence of $C$ suggests
that slight inhomogeneities in the sample may have pronounced
effects on $C$. Besides being difficult to quantify and
characterize, inhomogeneities cannot be included in a VCA
calculation, as disorder is averaged out in the VCA. The second
reason is that the measured $C$ implies that the magnetic entropy
$S$ at $T_N$ is remarkably reduced with respect to what expected
($S=k_{\rm B}(0.25 \ln(4)+0.75 \ln(3))$ per actinide ion), while the
measured $S$ is lower than $k_B \ln(2)$ per actinide ion (entropy
reduction is observed in pure $\mathrm{UO_2}$ and $\mathrm{NpO_2}$ as well, although
the reduction is smaller). Thus, it is {\it a priori} impossible
for any mean-field model such as the VCA to reproduce the measured
$C$ quantitatively. A qualitative feature of $C$ whose origin is
understandable within the VCA is the presence of slight
oscillations below $T_N$. These seem to reflect the onset of
contributions to the free energy from the ''delayed" ordering of
Np ions. These contribute little to the main peak of $C$ at $T_N$,
and produce features below $T_N$ when their ordering begins to
become appreciable.

The VCA provides a qualitatively satisfactory theoretical
framework for $\mathrm{U_{0.75}Np_{0.25}O_{2}}$, yet there is one point
(besides the entropy reduction problem) which is difficult to
understand. This is the low value of the saturation ordered moment
of Np ions. There appears to be no way to obtain a moment as low
as 0.6$\mu_{\rm B}$ with the observed 3-$\vec{k}$ structure.\cite{Caciuffo-EL}
This problem would remain even if theoretical approaches different
from the VCA were used, as it is associated with the structure of
the single-ion CF states of Np. These appear to carry a moment
along $\langle111\rangle$ always much larger than 0.6$\mu_{\rm B}$, no matter
whether the CF is varied or whether quadrupole interactions are
included. In particular, since the dipole and quadrupole order
parameters appearing in the 3-$\vec{k}$ structure commute,
the bare CF value of the moment cannot be reduced by quadrupolar
interactions. This would not be true if the structure was
1-$\vec{k}$.

The experiments on the mixed oxides have shown that at the 25\% Np doping ($x = 0.25$) the transverse nature of the $3-\vec{k}$ transition predominates, despite the fact that in pure $\mathrm{NpO_{2}}$ the ordering of the quadrupoles is longitudinal $3-\vec{k}$. Although the agreement in the top frame of Fig. 6 is not outstanding, it must be remembered that these experiments were performed on an off-specular reflection [the (112) from a (111) face] and this creates more difficulties experimentally than when a specular reflection is examined, as was the case in pure $\mathrm{NpO_{2}}$.\cite{Paixao02} The crucial aspect is that the maximum and minimum for azimuthal scans agree with the calculations for a transverse $3-\vec{k}$ quadrupolar ordering and completely disagree for the longitudinal model. This change across the solid solutions, from transverse in $\mathrm{UO_{2}}$ to longitudinal in $\mathrm{NpO_{2}}$ also gives a possible mechanism for the unusual effects found in the range $0.50 < x < 0.75$ [Refs. \onlinecite{Caciuffo-EL,Caciuffo-JPL,Tabuteau-SSC}] as with increasing Np concentration increasing frustration can be expected between these competing quadrupolar interactions. It will therefore be particularly interesting to try and observe the quadrupolar ordering in the compositions with higher Np content.

Finally, with these observations of the quadrupolar ordering in the mixed oxides we note in each case the Lorentzian squared shape of the resonances and their displacement to lower energy by about 2~eV (see Figs. 4 and 5). They arise from the $F^{[2]}$ term in the RXS cross section\cite{Hannon,Hill} and have also been observed in the work we reported on high-order terms in multi$-\vec{k}$ structures.\cite{PRB} The strong similarity between these signals and those found\cite{Paixao02} in pure $\mathrm{NpO_2}$ suggests that in $\mathrm{NpO_2}$ as well the principle RXS signal arises from a quadrupole signal, and not a more complicated process, as suggested recently.

\begin{acknowledgments}
The authors wish to thank Matt Longfield for his help in the early part of these experiments. SBW and PJ would like
to thank the European Commission for support in the frame of the `Training and
Mobility of Researchers' programme. Part of this work was made possible thanks to the support of the European Community-Access to Research Infrastructures action of the Improving Human Potential Programme (IHP) in allowing access to the Actinide User Laboratory at the Institute for Transuranium-Karlsruhe under the contract HPRIÐCT-2001Ð00118.
\end{acknowledgments}

\appendix*
\section{Azimuthal Dependence}
\label{sec:appendix}
In a cubic 3-$\vec{k}$ structure the complete star of $\vec{k}=\langle k 0 0\rangle$ is
simultaneously present in each volume element. Therefore there is only
one k-domain. However, several orientations of the Fourier components of the
magnetic moment relative to the corresponding wave vector are
possible. In particular, there are one longitudinal and two transverse
directions.

The longitudinal orientation is unique, so for this structure there is
only one S-domain. This would be the structure of $\mathrm{NpO_2}$
\emph{if there was a magnetic dipole moment}. In the transverse case, there are two possible orientations which may
be converted into each other by applying crystallographic symmetries.
This leads to 2 equivalent but distinct S-domains. This is the case in
$\mathrm{UO_2}$.

Below, we calculate the quadrupolar ($F^{[2]}$ term of the cross
section) scattering from these 3 constructions.

We assume the crystal structure is FCC with 4 atoms per unit cell. For each atom in turn we can calculate the moment direction at this position for the longitudinal and transverse domains (See Figure~\ref{Fig1}). The resulting directions are given in 
Table~I.

\begin{table}
\begin{ruledtabular}
  \begin{tabular}{cccc}
    Atom & Long. & Trans. Domain A & Trans. Domain B \\ \hline
    \miller{0}{0}{0} & $\vec{\mu}_{1} = \miller{1}{1}{1}$ & $\vec{\mu}_{1} = \miller{1}{1}{1}$ & $\vec{\mu}_{1} = \miller{1}{1}{1}$ \\
    \miller{\frac{1}{2}}{\frac{1}{2}}{0} & $\vec{\mu}_{2} = \miller{\overline{1}}{\overline{1}}{1}$ & $\vec{\mu}_{2} = \miller{1}{\overline{1}}{\overline{1}}$ & $\vec{\mu}_{1} = \miller{\overline{1}}{1}{\overline{1}}$ \\
    \miller{\frac{1}{2}}{0}{\frac{1}{2}} &  $\vec{\mu}_{3} = \miller{\overline{1}}{1}{\overline{1}}$ & $\vec{\mu}_{3} = \miller{\overline{1}}{\overline{1}}{1}$ & $\vec{\mu}_{3} = \miller{1}{\overline{1}}{\overline{1}}$ \\
    \miller{0}{\frac{1}{2}}{\frac{1}{2}} & $\vec{\mu}_{4} = \miller{1}{\overline{1}}{\overline{1}}$ & $\vec{\mu}_{4} = \miller{\overline{1}}{1}{\overline{1}}$ & $\vec{\mu}_{4} = \miller{\overline{1}}{\overline{1}}{1}$ \\
  \end{tabular}
 \end{ruledtabular}
 \label{tab:moment-directions}
 \caption{Moment directions on the four atoms of the unit cell for the single longitudinal and two transverse S-domains}
\end{table}

To form the scattering tensor for quadrupole scattering we first evaluate the quadrupole
operator $\vec{Q}_{ij}$ which in cartesian coordinates is given by
\begin{equation}
  \vec{Q}_{ij} = \mu_{i} \mu_{j} 
  - \frac{1}{3}\delta_{ij}\sum_k (\mu_{k} \mu_{k}) 
\end{equation}
This matrix is symmetric and traceless. For all of the above
structures $\mu_i^2 = 1, \forall i$, which leads to the general form
\begin{equation}
  \vec{Q} =\left ( \begin{array}{ccc}
    0 & \mu_{x}\mu_{y} & \mu_{x}\mu_{z} \\
    \mu_{x}\mu_{y} & 0 & \mu_{y}\mu_{z} \\
    \mu_{x}\mu_{z} & \mu_{y}\mu_{z} & 0  \end{array} \right )
\end{equation}

\begin{table}
\begin{ruledtabular}
\begin{tabular}{cccc}
Atom & Long. & Trans. Domain A & Trans. Domain B \\
\hline
 \miller{0}{0}{0} & $\left ( \begin{array}{ccc} 
0 & 1 & 1 \\ 1 & 0 & 1 \\ 1 & 1 & 0\\
\end{array} \right ) $ & $
\left ( \begin{array}{ccc} 
0 & 1 & 1 \\ 1 & 0 & 1 \\ 1 & 1 & 0\\
\end{array} \right ) $ & $
 \left ( \begin{array}{ccc} 
0 & 1 & 1 \\ 1 & 0 & 1 \\ 1 & 1 & 0\\
\end{array} \right ) $ \\ \\
 \miller{\frac{1}{2}}{\frac{1}{2}}{0} & $  \left ( \begin{array}{ccc} 
0 & 1 & \overline{1} \\ 1 & 0 &  \overline{1} \\ \overline{1} &  \overline{1} & 0\\
\end{array} \right ) $ & $
\left ( \begin{array}{ccc} 
0 & \overline{1} & \overline{1} \\ \overline{1} & 0 & 1 \\ \overline{1} & 1 & 0\\
\end{array} \right ) $ & $
 \left ( \begin{array}{ccc} 
0 & \overline{1} & 1 \\ \overline{1} & 0 & \overline{1} \\ 1 & \overline{1} & 0\\
\end{array} \right ) $ \\ \\
\miller{\frac{1}{2}}{0}{\frac{1}{2}} & $ \left ( \begin{array}{ccc} 
0 & \overline{1} & 1 \\ \overline{1} & 0 & \overline{1} \\ 1 & \overline{1} & 0\\
\end{array} \right ) $ & $
 \left ( \begin{array}{ccc} 
0 & 1 & \overline{1} \\ 1 & 0 &  \overline{1} \\ \overline{1} &  \overline{1} & 0\\
\end{array} \right ) $ & $
 \left ( \begin{array}{ccc} 
0 & \overline{1} &  \overline{1} \\ \overline{1} & 0 & 1 \\  \overline{1} & 1 & 0\\
\end{array} \right ) $ \\ \\
\miller{0}{\frac{1}{2}}{\frac{1}{2}} & $\left ( \begin{array}{ccc} 
0 & \overline{1} & \overline{1} \\ \overline{1} & 0 & 1 \\ \overline{1} & 1 & 0\\
\end{array} \right ) $ & $
\left ( \begin{array}{ccc} 
0 & \overline{1} & 1 \\ \overline{1} & 0 &  \overline{1} \\ 1 &  \overline{1} & 0\\
\end{array} \right ) $ & $
 \left ( \begin{array}{ccc} 
0 & 1 &  \overline{1} \\ 1 & 0 & \overline{1} \\  \overline{1} & \overline{1} & 0\\
\end{array} \right ) $ \\ 
\end{tabular}
\end{ruledtabular}
\label{tab:tensors}
\caption{Scattering tensors for the four atoms in the FCC unit cell for the longitudinal and transverse cases.}
\end{table}

Evaluating this for the 3 S-domains from the moment directions given in Table~I gives scattering tensors for each atom in the unit cell given by Table~II for the quadrupole moment.

From these the structure factor can be created as 
\begin{equation}
\vec{f} = \sum_{n} \vec{T}_{n} e^{\vec{q}\cdot\vec{r}},
\end{equation}
which for the Bragg reflection \miller{1}{1}{2} gives
\begin{eqnarray}
\vec{f}_{(112)} &=& \vec{T}_{1} + \vec{T}_{2} - \vec{T}_{3} - \vec{T}_{4}
\end{eqnarray}
yielding scattering tensors of the form
\begin{eqnarray}
\vec{f}^{L}_{(112)} = \left ( \begin{array}{ccc}
0 & 1 & 0 \\ 1 & 0 & 0 \\ 0 & 0 & 0 \\
\end{array}\right ) \\ \nonumber
\vec{f}^{A}_{(112)} = \left ( \begin{array}{ccc}
0 & 0 & 0 \\ 0 & 0 & 1 \\ 0 & 1 & 0 \\
\end{array}\right ) \\ \nonumber
\vec{f}^{B}_{(112)} = \left ( \begin{array}{ccc}
0 & 0 & 1 \\ 0 & 0 & 0 \\ 1 & 0 & 0 \\
\end{array}\right ) \nonumber
\end{eqnarray}

This can be evaluated to give a total structure factor of
\begin{equation}
\vec{F}^{[2]}_{(112)} = \vec{\epsilon}'\cdot\vec{f}_{(112)}\cdot\vec{\epsilon},
\label{eqn:Amplitude}
\end{equation}
where $\vec{\epsilon}$ and $\vec{\epsilon}'$ are the polarization
vectors of the incident and scattered beams, respectively. Here we
consider only $\sigma\rightarrow\sigma$ scattering, as the
$\sigma\rightarrow\pi$ channel is dominated by magnetic scattering
($F^{[1]}$) term.  

To evaluate this, first we construct a coordinate system of the crystal. We therefore, construct 3 unit vectors based on the reciprocal space vectors $\vec{q}$ (the scattering vector) and $\vec{k}_{\mathrm{azref}}$ (the azimuth reference vector), $\vec{\hat{c}}_{1}, \vec{\hat{c}}_{2}$ and $\vec{\hat{c}}_{3}$. These are related by,
\begin{eqnarray}
\vec{\hat{c}}_{3} & = & - \frac{\vec{q}}{|\vec{q}|}\nonumber\\
\vec{\hat{c}}_{2} & = & \vec{\hat{c}}_{3} \times \left ( \frac{\vec{k}_{\mathrm{azref}}}{|\vec{k}_{\mathrm{azref}}|} \right )\\
\vec{\hat{c}}_{1} & = & \vec{\hat{c}}_{2} \times \vec{\hat{c}}_{3}\nonumber.
\end{eqnarray}
We then do a rotation around $\vec{\hat{c}}_{3}$ to transform into the coordinate system of Blume and Gibbs.\cite{Gibbs} We therefore obtain 3 new unit vectors of the form,
\begin{eqnarray}
\vec{\hat{u}}_{3} & = & \vec{\hat{c}}_{3}\nonumber\\
\vec{\hat{u}}_{2} & = & \vec{\hat{c}}_{1}\sin\phi + \vec{\hat{c}}_{2}\cos\phi\\
\vec{\hat{u}}_{1} & = & \vec{\hat{c}}_{1}\cos\phi - \vec{\hat{c}}_{2}\sin\phi\nonumber.
\end{eqnarray}
which again are linear combinations of the reciprocal space vectors $\vec{q}$, $\vec{k}_{\mathrm{azref}}$ and the cross product of these two.
We can then define the polarization vectors for the incident and exit beams in the new coordinate system. For $\sigma$ polarization this is simply equal to the unit vector $\vec{\hat{u}}_{2}$, i.e. the polarization vector is normal to the scattering plane (given by the cross product of the scattering vector and the azimuthal reference vector). For $\pi$ polarization the Bragg angle of the reflection has to be taken into consideration. This is included here for completeness but is not used for these calculations.
We therefore have the polarization vectors in the rotated coordinate system given by 
\begin{eqnarray}
\vec{\hat{\epsilon}}_{\sigma} &=& \vec{\hat{\epsilon}}_{\sigma}'  =  \vec{\hat{u}}_{2}\\
\vec{\hat{\epsilon}}_{\pi} & = & \vec{\hat{u}}_{1}\sin\theta_{B} - \vec{\hat{u}}_{3}\cos\theta_{B}\\
\vec{\hat{\epsilon}}_{\pi}' & = & -\vec{\hat{u}}_{1}\sin\theta_{B} - \vec{\hat{u}}_{3}\cos\theta_{B}
\end{eqnarray}
Given these vectors equation~\ref{eqn:Amplitude} can be evaluated to give the amplitude of the scattering for each required $\phi$ value. 

\end{document}